\documentclass[10pt]{iopart}

\usepackage{bm}
\usepackage{graphicx}
\usepackage{color,soul}
\usepackage[T1]{fontenc}
\usepackage{ae,aecompl}

\begin{document}

\title[Comment on ``If it's pinched it's a memristor'']{Comment on ``If it's pinched it's a memristor'' by
L.~Chua [Semicond. Sci. Technol. {\bf 29}, 104001  (2014)]}
\author{Y.~V.~Pershin}
\ead{pershin@physics.sc.edu}
\address{Department of Physics and Astronomy, University of South Carolina, Columbia, South Carolina
29208, USA}

\author{M.~Di Ventra}
\ead{diventra@physics.ucsd.edu}
\address{Department of Physics, University of California San Diego, La Jolla, California 92093, USA}

\vspace{10pt}
\begin{indented}
\item[]May 2019
\end{indented}

\begin{abstract}
In his paper ``If it's pinched it's a memristor'' [Semicond. Sci. Technol. {\bf 29}, 104001  (2014)] L. Chua claims to extend the notion of memristor to all two-terminal resistive devices that show a hysteresis loop pinched at the origin. He also states that memcapacitors and meminductors can be defined by a trivial replacement of symbols in the memristor relations, and, therefore, there should be a correspondence between the hysteresis curves of different types of memory elements. This leads the author to the {\it erroneous conclusion} that charge-voltage curves of {\it any} memcapacitive devices should be pinched at the origin. The purpose of this Comment is to correct the wrong statements in Chua's paper, as well as to highlight some other inconsistencies in his reasoning. We also provide  experimental evidence of a memcapacitive device showing non-pinched hysteresis.
\end{abstract}


Although resistive devices and systems with memory were well known in the literature (both experimental and theoretical) well before the 70's, the name ``memristive elements'' and their formal definition introduced in 1976 by Chua and Kang~\cite{chua76a} are now widely used to describe various types of resistance switching memories and other kinds of memory systems and devices (see e.g., Ref.~\cite{pershin11a} for an extended review). At the same time, the ``ideal memristor'' (in the sense of its original definition~\cite{chua71a}) still remains an elusive/idealized concept, leading
many researchers to raise serious concerns about its actual existence as a
physically-realizable device~\cite{mouttet2012memresistors,meuffels2012fundamental,di2013physical,vongehr2015missing,sundqvist2017memristor}. In fact, it is doubtful that
any experimental system would pass the memristor test we have recently proposed in Ref.~\cite{pershin18a}, which would differentiate between an ideal memristor and
the more general, and physically-valid concept of memristive element.

In his paper~\cite{Chua_2014}, Chua has further pushed the idea of the ``universality'' of memristors, and even went a step further by claiming that memcapacitive and meminductive~\cite{diventra09a} devices and systems can be defined by a trivial replacement of symbols in the memristor relations. Therefore, he claims there should be a correspondence between the hysteresis curves of different types of memory elements.

In this Comment we point out several errors and misleading
statements in Ref.~\cite{Chua_2014}. In fact, we demonstrate that it is really {\it not} necessary for the different memelements to show pinched characteristics to be valid memory
elements. It is enough for them to have {\it i}) a specific type of response (resistive, capacitive, or
inductive), and {\it ii}) a memory component. Nothing else is required of them.

Our perspective rests on the {\it fact} that memristive, memcapacitive, and meminductive devices and systems are simply generalizations of the traditional resistors, capacitors, and inductors to the case of memory response~\cite{di2013physical}. This is not an assumption or a mathematical definition. It is a matter of {\it physical reality}: {\it any} physical system shows some degree of memory in its response to external perturbations~\cite{kubo1957statistical}, whether that memory is easy to
detect or not. Therefore, since theoretical models have to correspond to an actual physical reality to be valid models (or else what is done is at best mathematics, at worst pseudo-science), the basic properties of memory device models (whether resistive, capacitive or inductive) should reflect those of the corresponding experimental systems and devices. With these important preliminaries in mind, let us then highlight the errors and misleading statements in Ref.~\cite{Chua_2014}.

{\it First point:} the hysteresis curves pinched at the origin {\it need not} result from non-divergent memory resistances or conductances. In fact, it is straightforward to show that the pinching at the origin is still possible with some divergent resistance (leading to an {\it insulating} state) and conductance (leading to a {\it superconducting} state). Why these physically realizable situations should not be part of a well-defined memory element is a mystery to us.

Consider, for instance, a current-controlled memristive system described by
\begin{eqnarray}
  V &=& R_M(x,I)I\,,   \label{eq:1a}\\
  \dot{x} &=& f(x,I) \,, \label{eq:1b}
\end{eqnarray}
with $R_M(x,I)=g(x)/\sqrt{|I|}$. Here, $V$ and $I$ are the voltage across and current through the memristive system, respectively, $R_M(x,I)$ is the memory resistance, $x$ is the internal state variable, $f(x,I)$ is the function describing the evolution of $x$, and $g(x)\geq 0$ is a bounded function of $x$. It is evident that as $I$ goes to zero
\begin{equation}\label{eq:2}
  V=\lim\limits_{I\rightarrow 0} R_M(x,I)I=0,
\end{equation}
while
\begin{equation}\label{eq:3}
  \lim\limits_{I\rightarrow 0} R_M(x,I)=\infty,
\end{equation}
for $g(x)> 0$, and yet the above model is a valid memristive element describing a transition to an insulating state.

In fact, we emphasize that Eq.~(\ref{eq:2}) is a superior criterion for the $I-V$ curve passing through the origin compared to the criterion $R_M(x,0)\neq \infty$ introduced in Ref. \cite{Chua_2014}, which would exclude physically plausible situations.

{\it Second point:} the author of Ref.~\cite{Chua_2014} criticizes our prior work~\cite{martinez09a} by stating that ``a memcapacitor described by a non-pinched hysteresis loop in the $q$ versus $V$ plane is erroneous. Their error can be traced to the associated capacitance tending to infinity at the origin''~\cite{Chua_2014}.

Chua here reaches these erroneous conclusions due to a clear mistake in his reasoning. His idea that memcapacitive systems and their properties can be derived/formulated by a trivial replacement of ``symbols'' in the memristive relations does not hold since the physics of memristive and memcapacitive elements is too different. In particular, the memory properties of resistors are often based on electronic and ionic transport~\cite{pershin11a}. On the other hand, memcapacitors can support memory by, for instance, micro-electro-mechanical effects~\cite{pershin11c} (e.g., changes of the relative position of their plates) or time delay in the relative permittivity of the medium in between the capacitor plates~\cite{pershin11a}. Both cases are physically possible and both lead to memory effects. In fact, even the basic properties of the usual resistors and capacitors, such as the passivity of resistors and the reactance of capacitors, do not allow a transformation from one element to the other by a simple ``replacement of symbols''.

In the case of our specific memcapacitor realization~\cite{martinez09a} criticized in Ref.~\cite{Chua_2014}, the memory effect is caused by the delayed/nonlinear response to the applied bias of a multilayer medium positioned between the capacitor plates. This medium shifts the charging/discharging $q-V$ curves with respect to each other and away from the origin. Clearly, this is not even remotely similar to what occurs in memristive systems.
\begin{figure}[t]
  \centering
  (a)\includegraphics[width=60mm]{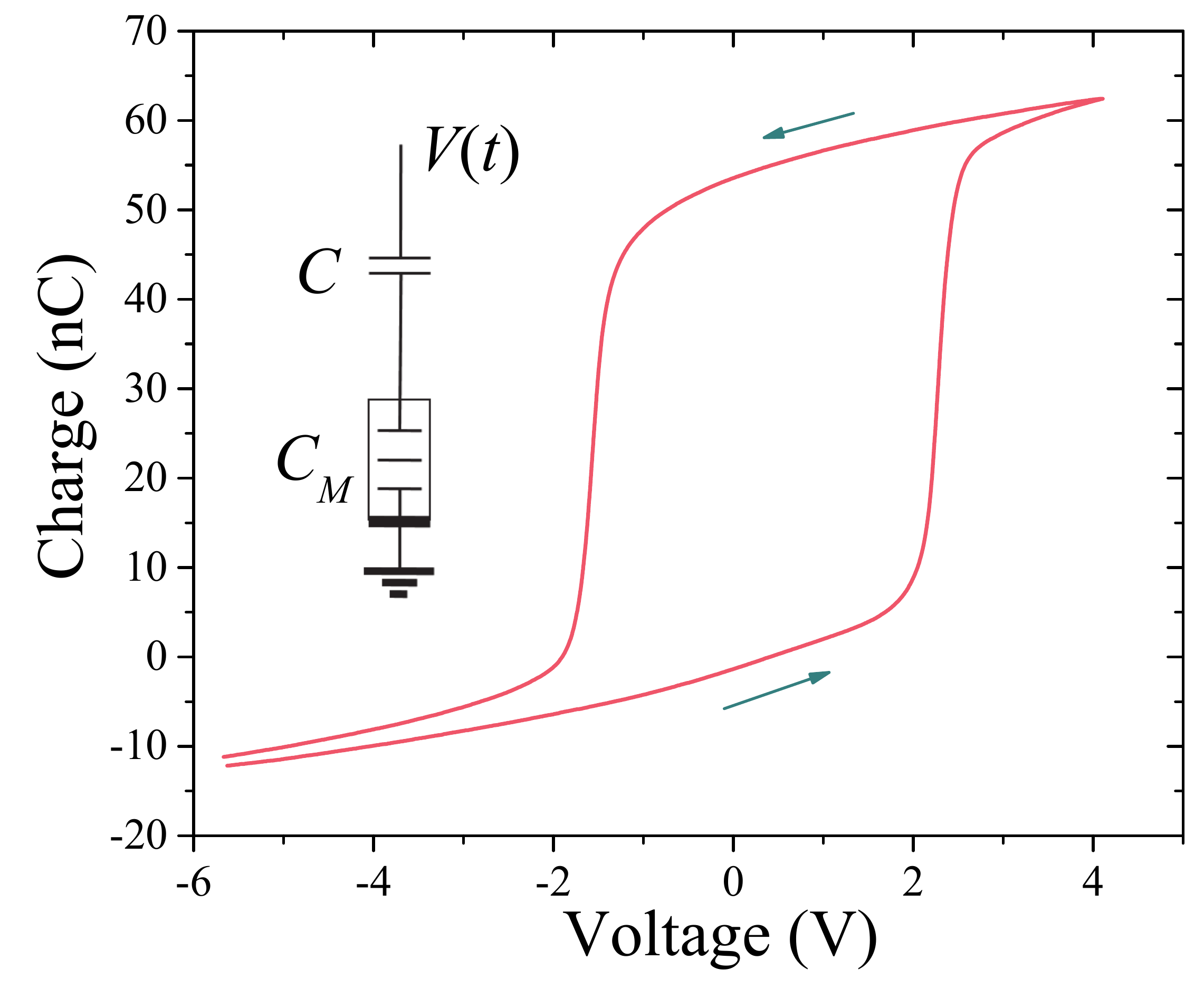} (b)\includegraphics[width=60mm]{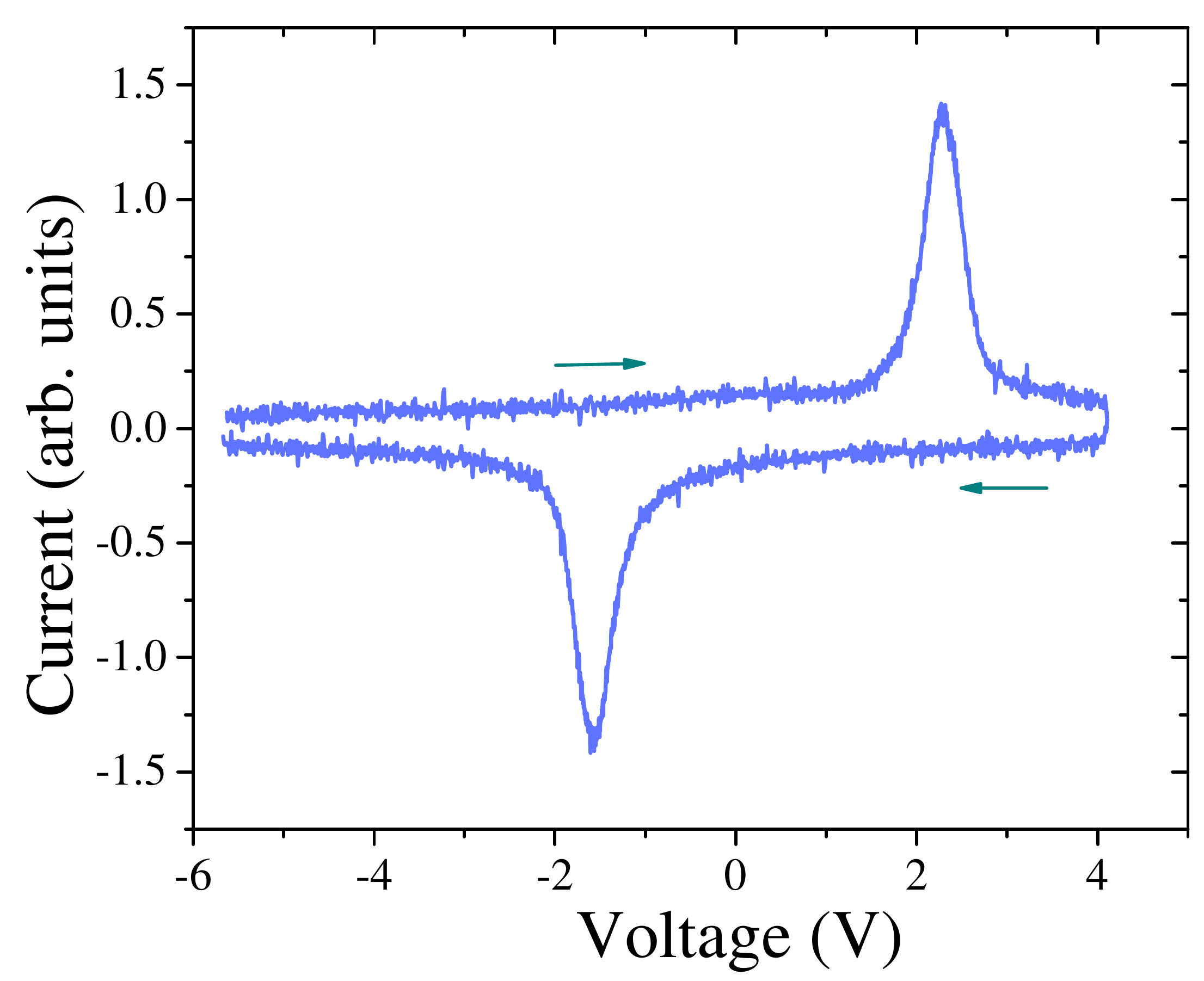}
  \caption{Experimental electric response of a ferroelectric memcapacitor. (a) $q-V$ curve, and (b) $I-V$ curve. The inset of (a) shows the schematic of the  capacitor-ferroelectric capacitor circuit used in our measurements. A triangular voltage waveform was used in the measurements. In this figure, the
  	measured charge, $q$,  and current, $I$, are plotted as functions of the voltage across the memcapacitor. The measurement was performed using a standard capacitor of $C=33$~nF, and a lead zirconate titanate (PZT) ferroelectric capacitor with film thickness of 255~nm and plate area of $10^5$ $\mu$m$^2$. }\label{fig1}
\end{figure}

To further support our arguments, consider the experimentally-obtained $q-V$ curve of a ferroelectric capacitor as reported in Fig. \ref{fig1}(a). Previously we pointed out that ferroelectric capacitors are a kind of memcapacitors~\cite{pershin11a}, and there is no doubt about such classification~\cite{flak2014solid}. We have performed experimental measurements of the $q-V$ curves of a lead zirconate titanate (PZT) ferroelectric capacitor using a simple capacitor-ferroelectric capacitor circuit shown in the inset of Fig. \ref{fig1}(a). The circuit was driven by a triangular voltage waveform, and the charge on the ferroelectric capacitor was measured from the voltage drop across the standard capacitor. Fig.~\ref{fig1}(a) clearly shows that the $q-V$ curve of the ferroelectric memcapacitor is {\it not pinched}. The $I-V$ curve obtained by the differentiation of the experimentally measured $q(t)$ exhibits two peaks typical of ferroelectric capacitors~\cite{Schenk14a} (see Fig.~\ref{fig1}(b)).

{\it Third point:} by referring again to our work~\cite{martinez09a} Chua writes in Ref.~\cite{Chua_2014} ``It also follows that their rather intimidating quantum-mechanical arguments were incorrectly applied''. Chua does not seem to realize that our work~\cite{martinez09a} is based entirely on classical electrodynamics, except for the use of the Simmons formula for the tunneling current~\cite{Simmons1963-1}. We are surprised that the Simmons current expression was not recognized by the author of Ref.~\cite{Chua_2014}. There should be nothing ``intimidating'' about using classical electrodynamics and standard formulas for the tunneling current: it should be standard knowledge of any electrical engineer.

{\it Fourth point:} the word ``pinched'' is {\it not general} enough to represent all kinds of memristive hysteresis.
Semantically, a ``pinched loop'' is a loop obtained by the action of pinching. In a pinched loop,
the ascending and descending curves  are tangent to each other at the pinched point (see, e.g., Fig.~2(b) in~\cite{Chua_2014}).
To obtain the self-crossing hysteresis curves, which are the most common ones (see Fig.~2(a) in~\cite{Chua_2014}), the hysteresis loop needs to be {\it twisted} not pinched.

{\it Final point:} although one can loosely use the word ``memristor'' to indicate {\it any} memristive elements (and we have done so sometimes as well), it is
very important to remember that an actual ``ideal memristor'' as originally defined in~\cite{chua71a} {\it has not been found yet}. By using the same name
for all resistive memories seems only an attempt to distract from this important fact. As we have already argued in our Ref.~\cite{pershin18a}, before claiming that such a
hypothetical device had been found, researchers should submit their devices to the test we have proposed in~\cite{pershin18a}, which clearly and {\it unambiguously} distinguishes between an ideal memristor and actual, physically possible and well-documented memristive devices.

To summarize, the present Comment corrects wrong and misleading statements and concepts presented by Chua in Ref.~\cite{Chua_2014}.
In particular, we have shown that {\it i})
the hysteresis curves pinched at the origin {\it need not} result from non-divergent memory resistances or conductances, and  divergent response functions {\it
	do not} preclude different physical systems to be valid memory
elements, {\it ii}) a trivial replacement of symbols in the memristor relations leads to {\it incorrect conclusions} regarding the hysteresis curves of memcapacitors and meminductors, {\it iii}) memcapacitors with non-pinched hysteresis {\it do exist}, and we have even provided experimental evidence of one such case.

In other words, physically realizable memristive, memcapacitive and minductive elements are simply resistors, capacitors and inductors whose memory can
be detected experimentally. Whether their constitutive relations are pinched at the origin or not {\it is irrelevant}. A lesson can be drawn
from all this: over-reliance on mathematical definitions to describe physical phenomena can easily lead to the wrong conclusions and away from
physical reality.

\section*{References}

\bibliographystyle{IEEEtran}
\bibliography{memcapacitor}

\end{document}